\documentclass[aps,twocolumn,showpacs,superscriptaddress,citeautoscript,prl,floatfix, 10pt]{revtex4-2}
\usepackage{ulem}
\usepackage{bm}
\usepackage{graphicx}
\usepackage{amsmath,amssymb}
\usepackage{amsfonts}
\usepackage{color}
\usepackage{siunitx}
\usepackage{hyperref}
\usepackage{cleveref}

\definecolor{Red}{rgb}{0.9,0.3,0.3}

\begin{document}

\title{Topological bound states in the continuum}

\author{Dunkan Mart\'{\i}nez}
\email[Corresponding author: ]{dunmar01@ucm.es}
\affiliation{GISC, Departamento de F\'{\i}sica de Materiales, Universidad Complutense, E-28040 Madrid, Spain}

\author{Rodrigo P. A. Lima}
\affiliation{GISC \& InterTEP, Departamento de F\'{\i}sica, 
Facultad de Ciencias Ambientales y Bioqu\'{\i}mica, Universidad de
Castilla-La Mancha, 45071 Toledo, Spain}	
\affiliation{GFTC, Instituto de F\'{\i}sica, Universidade Federal de Alagoas, Macei\'{o} AL 57072-970, Brazil}

\author{Alexander L\'{o}pez}
\affiliation{GISC, Departamento de F\'{\i}sica de Materiales, Universidad Complutense, E-28040 Madrid, Spain}
\affiliation{Escuela Superior Polit\'ecnica del Litoral, ESPOL, Departamento de F\'isica, Facultad de Ciencias Naturales y Matem\'aticas, Campus Gustavo Galindo
 Km. 30.5 Via Perimetral, P. O. Box 09-01-5863, Guayaquil, Ecuador}

\author{Francisco Dom\'{\i}nguez-Adame}
\affiliation{GISC, Departamento de F\'{\i}sica de Materiales, Universidad Complutense, E-28040 Madrid, Spain}

\author{Pedro A. Orellana}
\affiliation{Departamento de F\'{\i}sica, Universidad T\'{e}cnica Federico Santa Mar\'{\i}a, Casilla 110 V, Valpara\'{\i}so, Chile}

\begin{abstract}


Bound states in the continuum, originally proposed within the framework of quantum mechanics, have since been observed in a variety of physical contexts, including electromagnetism, acoustics, and optics. Of particular interest are those bound states in the continuum that are protected by continuous symmetries, as their stability makes them resistant to structural imperfections and material disorder. In this study, we demonstrate the existence of topologically protected bound states in the continuum by coupling a finite Su–Schrieffer–Heeger (SSH) chain to a metallic lead. These states are characterized by distinct features in the electrical response of the system, serving as a direct indication of their topological origin. The inherent robustness of such topologically protected states highlights their potential applications in fault-tolerant quantum information processing as well as in the design of advanced electronic and photonic devices.
\end{abstract}

\maketitle

\textit{Introduction}---Bound states in the continuum (BICs) are counterintuitive quantum states that, despite residing within the energy range of extended states, remain perfectly localized and decoupled from the continuum. The absence of coupling gives these states an extremely long lifetime. First introduced by von Neumann and Wigner in 1929 for quantum particles in tailored potentials~\cite{Neumann1929}, BICs have been identified across diverse physical platforms, including electromagnetism~\cite{Bulgakov2008, Martinica2008, Plotnik2011, Molina2012, Corrielli2013, Hsu2013, Weimann2013, Monticone2014, Yang2014, Zhen2014, Vicencio2015, Mukherjee2015, Rivera2016, Gomis-Bresco2017, Azzam2018, Cerjan2019, Azzam2021, Song2023, Pinto2024, Legon2025, Martinez2025}, acoustic~\cite{Lyapina2015, Huang2020, Huang2021, Deriy2022} and water waves~\cite{McIver1996, Linton2007, Euve2023a}.  These states can emerge via several distinct mechanisms such as symmetry protection~\cite{Plotnik2011, Legon2025, Martinez2025, Parker1966, Cumpsty1971, Evans1994}, separations of coordinate variables~\cite{Rivera2016, Nockel1992}, parameter tuning~\cite{Bulgakov2008, Hsu2013, Martinica2008, Weimann2013, Monticone2014, Lyapina2015, Huang2020, Huang2021} and inverse construction~\cite{Molina2012, Corrielli2013, McIver1996}. A key characteristic of BICs is their extremely high quality factor, which arises from their decoupling from the continuum spectrum states. Furthermore, in some cases, the quality factor can be finely tuned, making BICs attractive for use in applications such as narrow-band filters, highly sensitive sensors, and low-threshold lasers~\cite{Hsu2016, Kodigala2017, Hwang2021, Yang2021, Mohamed2022, Hwang2022, Liu2023}. However, this sensitivity can also be a double-edged sword: small perturbations or imperfections can hybridize BICs with the continuum, destroying their localized nature. 

On the other side, topological systems have attracted a lot of attention~\cite{Su1979, Hasan2010, Qi2010, Lu2014, Xue2022} due to their unique properties, such as the protection of topological states against specific classes of defects~\cite{Lu2014, Ma2019, Xue2022, Martinez2023}, as well as their inherent robustness for quantum computation technologies~\cite{Kitaev2003, Milman2003, Souza2007, Oxman2011, Johanson2012, Matoso2016}.  The recent convergence of these two concepts has given rise to a new paradigm: topological BICs (TBICs), which combine the spectral isolation of BICs with the disorder-resilience of topological states~\cite{Yang2013, Xiao2017, Takeichi2019, Cerjan2020, Benalcazar2020, Li2020, Rivero2023, Liu2023b, Wang2023, Wang2024, Dong2025}.

In this work, we uncover TBICs in electron systems that arise from the coupling between a topological chain and a conventional metallic contact. Specifically, we study the effect of attaching a finite Su-Schrieffer-Heeger (SSH) chain~\cite{Su1979} by one of its ends to a metallic lead. Despite the simplicity of the setup, we find that the SSH spectrum leaves a clear fingerprint on the lead’s transmission profile. Most remarkably, a topological BIC emerges at zero energy: A perfectly confined state within the continuum, protected by the SSH chain’s topology rather than the conventional parameter fine-tuning. This setup offers a minimal but robust platform for realizing TBICs in quantum systems.

\begin{figure}[ht]
    \centering
    \includegraphics[width=\linewidth]{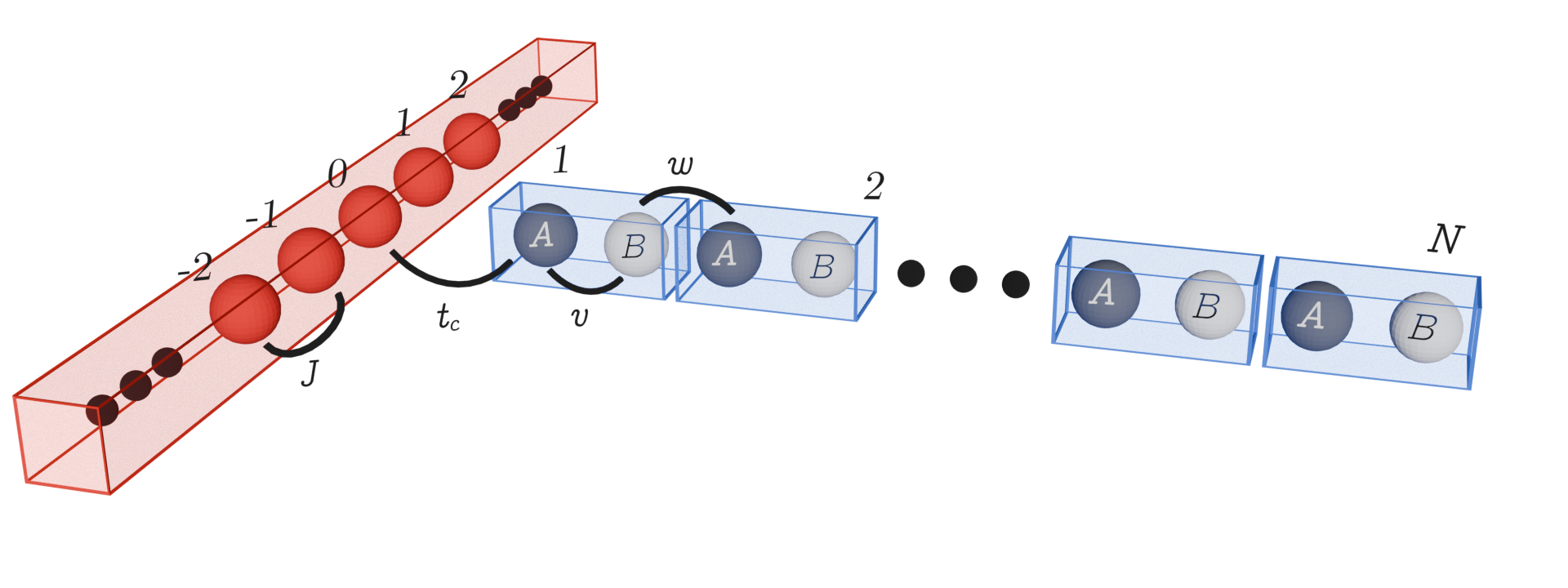}
    \caption{Sketch of the system. The red bar represents the infinite lead while the blue boxes show the dimer cells of the SSH chain with $N$ cells. The spheres within the boxes represent the sites of the aforementioned elements: Red spheres correspond to the sites of the lead while dark (light) spheres indicate sites in the sublattice~A~(B). Black curved lines indicate the hopping considered in our model.}
    \label{fig: 1}
\end{figure}

\textit{Model}---The system under study consists of a SSH chain of $N$ dimer cells ($2N$ sites), coupled to a lead at one of its ends, as shown schematically in Fig.~\ref{fig: 1}.
This arrangement enables the detection of SSH chain states as anti-resonances in the conductance of the lead. The system under study can be conveniently described by a tight-binding Hamiltonian, written as $\mathcal{H} = \mathcal{H}_{ch} + \mathcal{H}_l + \mathcal{H}_{c}$. Here $\mathcal{H}_{ch}$ is the Hamiltonian of the isolated SSH chain, $\mathcal{H}_l$ denotes the Hamiltonian of the isolated lead and $\mathcal{H}_c$ stands for the coupling between the lead and the SSH chain. To be specific, these terms are defined as
%
%
\begin{align}
    \mathcal{H}_{ch} =&  -v\sum_{\ell=1}^N  (a_\ell^\dagger b_\ell^{} + \mathrm{h.c.}) - w\sum_{\ell=1}^{N-1} (b_{\ell}^\dagger a_{\ell+1}^{} + \mathrm{h.c.})\ ,\nonumber  \\
    \mathcal{H}_l =& -J \sum_j (c_j^\dagger c_{j+1}^{} + \mathrm{h.c.})\ , \nonumber \\
    \mathcal{H}_c =& -t_c c_0^\dagger a_1^{} + \mathrm{h.c.}
\end{align}
%
%
Here, $c_j$ is the annihilation operator of an electron at site $j$ of the lead while $a_{\ell}$ ($b_{\ell}$) is the annihilation operator of the SSH chain at cell $\ell$ and sublattice $A$ ($B$). $J$ is the hopping parameter in the lead, $v$ is the intracell hopping in the chain while $w$ is the intercell hopping, and $t_c$ is the coupling parameter between the lead and the chain (see Fig.~\ref{fig: 1}). We have assumed that all site energies are identical and set them to zero, without loss of generality.

Since the coupling between the lead and the chain is local, we can assume that the conductance of the system can be calculated \sout{by} using the Landauer–Büttiker formalism in the ballistic regime as follows
\begin{equation}
    G(E) = \frac{\text{e}^2}{\hbar}\, \tau(E)\ ,
\end{equation}
where $\tau$ is the transmission coefficient at Fermi energy $E$ and $-$e is the electron charge. To determine the transmission, we write the wave function for each subsystem as follows
\begin{subequations}
    \begin{align}
        \phi_j = & \left\{ \begin{matrix} e^{i k j} + \hat{r} e^{-i k j}\ , & \text{if $j < 0$\ , } \\ \hat{t} e^{i k j}\ , & \text{otherwise}\ , \end{matrix} \right. \\
        \psi_{\ell} = & \begin{pmatrix}  \psi_{\ell}^A \\ \psi_{\ell}^B \end{pmatrix} = \alpha \begin{pmatrix}  1 \\ e^{i\theta} \end{pmatrix} e^{iq\ell} + \beta \begin{pmatrix}  1 \\ e^{-i\theta} \end{pmatrix} e^{-iq\ell}\ ,
    \end{align} 
\end{subequations}
where $\phi_j$ is the wave function at the site $j$ of the lead, while $\psi_\ell$ corresponds to the wave function in the cell $\ell$ of the chain. In addition, $\hat{r}=\hat{r}(E)$ and $\hat{t}=\hat{t}(E)$ are the energy-dependent reflection and transmission amplitudes, respectively, while $\alpha=\alpha(E)$ and $\beta=\beta(E)$ are the amplitude coefficients of the wave function inside the chain. Here we have taken into account that the difference between the wave function of the two sublattices in the same cell is just a complex phase  $\exp(i \theta) = -E/[v + w\exp(-iq)]$~\cite{Bissonnette2023}. In addition, the wavenumbers $k$ and $q$ can be found from the dispersion relation of the lead and chain, respectively, namely $k = -\arccos(E/2J)$ and $q = \arccos\left[(E^2 - v^2 - w^2)/2vw\right]$. Thus, the boundary conditions for this problem can be written as
%
    \begin{align}
        E \phi_{-1} =& -J (\phi_0 + \phi_{-2})\ ,\nonumber\\
        E \phi_0 =& -t_c \psi_1^A -J(\phi_1 + \phi_{-1})\ ,\nonumber\\
        E \psi_1^A = & -v \psi_1^B - t_c\phi_0\ ,\nonumber\\
        E \psi_N^B =& -v\psi_N^A \ .
    \end{align}
%
After some lengthy but straightforward calculations (see the Supplementary Material), the transmission amplitude is found to be given by
\begin{equation}
    \hat{t}(E) = \frac{1}{1-i \Gamma\, \dfrac{\sin [q(E)N]}{\sin k(E) \sin [q(E)(N+1)-\theta(E)]}} \ ,
    \label{eq:transmission}
\end{equation}
where $\Gamma =  t_c^2/2Jw$ is an energy-independent parameter in the wide-band limit describing the effective hopping between the SSH chain and the lead. Once we know the expression for the transmission coefficient, we can find the values of the parameters $\alpha$ and $\beta$ to obtain the electron wave function in the chain $\psi_\ell$. This can be used to find the density of the states~(DOS) in the chain as $\rho^\mathrm{ch}(E) \sim \sum_{\ell \in \mathrm{chain}}|\psi_\ell(E)|^2$. 

\begin{figure}
    \centering
    \includegraphics[width=\linewidth]{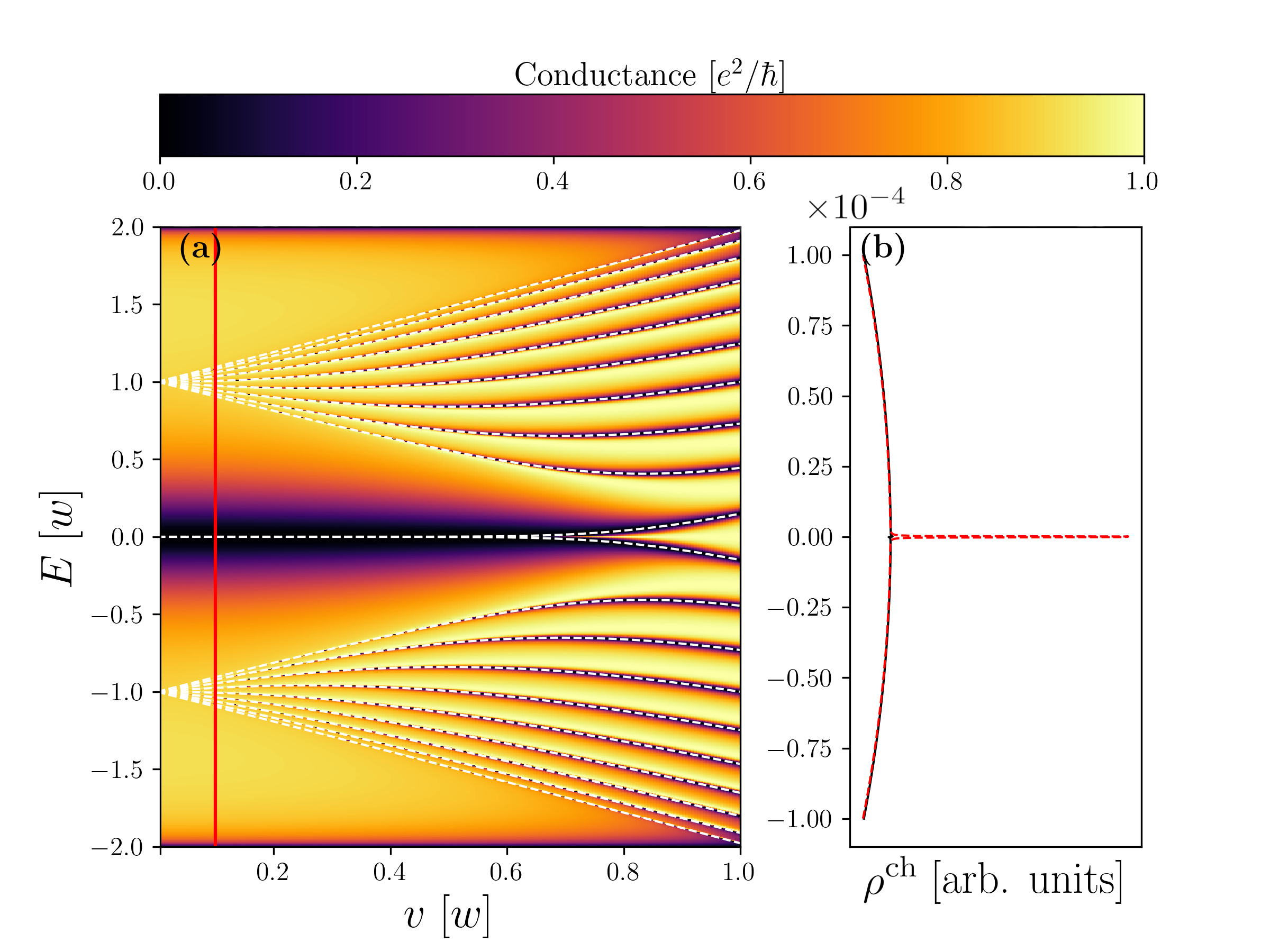}
    \caption{Conductance and DOS of a SSH chain of $N=10$ cells hanged to the lead, $\Gamma$ has been set to $0.3$ and $J$ to $1$. (a)~Conductance in units of $e^2/\hbar$, in white dashed lines the spectra of the SSH has been plotted. (b)~DOS in arbitrary units for $v=10^{-5}$ (solid black) and $v=0.1$ (dashed red) near $E=0$ (notice the scale factor).}
    \label{fig:Map+DOS}
\end{figure}

\textit{Topological BIC}--- From Eq.~\eqref{eq:transmission} we can see that the transmission at the Fermi energy can be tuned by the parameters $v$, which effectively controls the occurrence of the topological state of the chain, and $\Gamma$. For the present study, we consider $w$ as the unit of energy. Therefore, the topological state arises if $v<w$. As we are dealing with a perfect lead with a SSH chain attached to it, we expect the transmission to be unity for all energies except the ones that match the energy spectra of the latter [see Fig.~\ref{fig:Map+DOS}(a)].

We see from the results shown in Fig.~\ref{fig:Map+DOS}(b) that the DOS remains almost independent of energy except near $E=0$, where a resonance appears, exhibiting the expected behavior of a TBIC. This TBIC comes from the exponential decay that characterizes the topological state in the SSH chain. When $v$ is close to zero, the exponential decay is very abrupt, causing the wave functions at either end to barely overlap. This fact allows us to combine both topological states in a symmetric and antisymmetric way, resulting in a state that is effectively disconnected from the lead. 

\begin{figure}[ht]
    \centering
 \includegraphics[width=\linewidth]{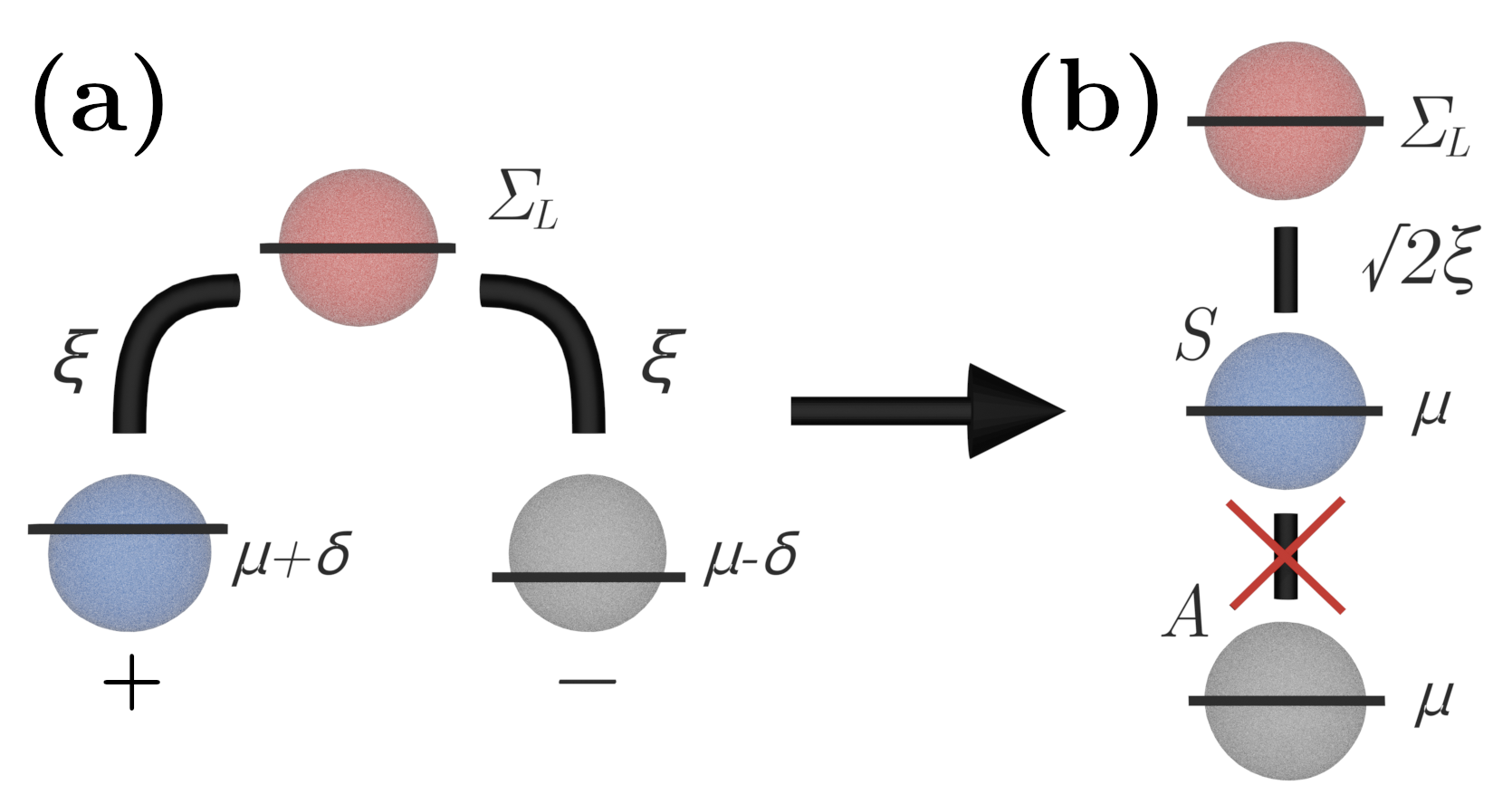}
    \caption{Sketch of the effective model that illustrates the formation of the TBIC. (a)~The entire chain is effectively replaced by a single site with complex energy  $\Sigma_L$ that is coupled with coupling constant $\xi$ to two topological states with energies $\mu\pm\delta$ (see main text for an explanation). (b)~After performing an unitary transformation, the antisymmetric state $A$ is decoupled from the rest of the states and it can be regarded as a TBIC.}
    \label{fig:sketch2}
\end{figure}

To explain the emergence of TBICs, we consider an effective model in which the influence of the infinite lead is embodied through a renormalized site with complex self energy $\Sigma_L$, coupled via an effective coupling constant $\xi$ to the two topological states of the SSH chain with energies $\mu \pm \delta$ (see Fig.~\ref{fig:sketch2}). The effective Hamiltonian reads
\begin{eqnarray} 
H_\mathrm{eff}&=& \Sigma_L|L\rangle\langle L|+(\mu+\delta) |\psi_+\rangle\langle \psi_+|
+(\mu-\delta) |\psi_-\rangle\langle \psi_-| 
\nonumber \\
&-&\xi|L\rangle\Big (\langle \psi_+|+\langle \psi_-|\Big)-\xi^*\Big( |\psi_+\rangle +|\psi_-\rangle\Big)\langle L|\ .
\end{eqnarray}
By performing a linear transformation to the symmetric and antisymmetric states, $|\psi_S\rangle=(1/\sqrt{2})(|\psi_+\rangle +|\psi_-\rangle)$, $|\psi_A\rangle=(1/\sqrt{2})(|\psi_+\rangle -|\psi_-\rangle)$ and taking the limit $\delta \to 0$, i.e., the topological regime, 
the Hamiltonian takes the form
\begin{eqnarray}
   H_\mathrm{eff}= \Sigma_L|L\rangle\langle L|+\mu |\psi_S\rangle\langle \psi_S|+\mu |\psi_A\rangle\langle \psi_A|\\ \nonumber-\sqrt{2}\xi|L\rangle \langle \psi_S|-\sqrt{2}\xi^*|\psi_S\rangle \langle L|\ . 
\end{eqnarray}
Therefore, the antisymmetric state is an eigenstate of the entire system and thus a TBIC, a bound state in the continuum protected by the system's topology. Figure \ref{fig:sketch2} shows a sketch of the effective model that illustrates the formation of the TBICs through the linear transformation described above.


%
%
%


%
%

\textit{Dynamics of the TBICs}--- The localized nature of the state can be assessed by the time-evolution of a wave packet initially localized at one of three representative sites within the SSH chain: The nearest site to the lead, where the wave function amplitude decays exponentially as it leaks into the lead, a second site within the bulk of the chain, where the dynamics consists of multiple reflections between the inner boundaries of the chain, producing Fabry–Perot–like interference. Finally, at the farthest site where the BIC is located, the wave function should remain localized and time-invariant. 

\begin{figure}
    \centering
 \includegraphics[width=\linewidth]{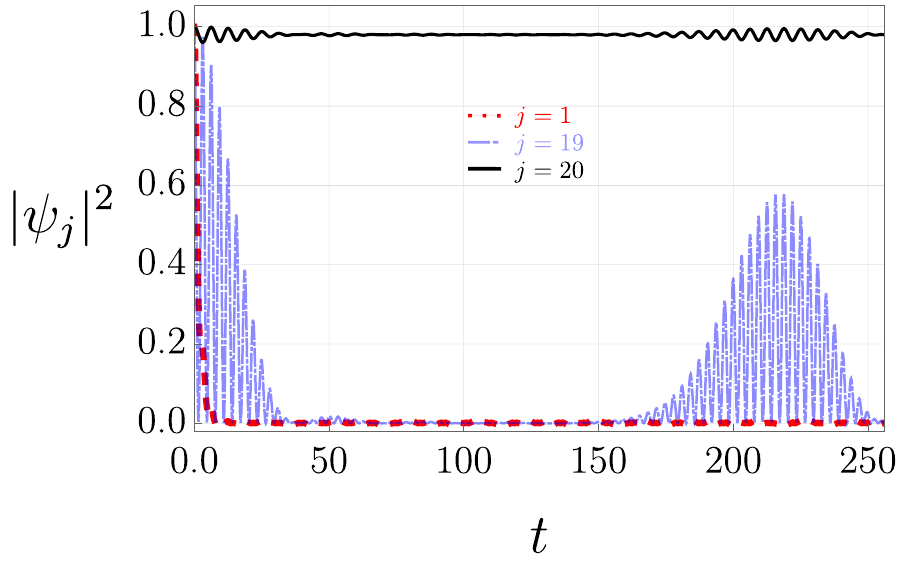}
    \caption{Return probability as a function of time of two topological states (dotted red and solid black curves) and one trivial state (dashed blue curve) on a SSH chain of $N=10$ dimer cells attached to the lead, with $\Gamma=0.3$, $J=1$ and $v=0.1$, for three specific sites: $j=1$ (dotted red line), $j=19$ (dashed blue line) and $j=20$ (solid black line). Time is expressed in units of $w^{-1}$.}
    \label{fig:rp1}
\end{figure}

Figure~\ref{fig:rp1} displays the time evolution for the three initial conditions described above. The red line corresponds to the site nearest to the lead, where the probability amplitude decays approximately exponentially as it couples to the continuum. The blue line represents a state initialized at the farthest boundary of the SSH bulk, which propagates along the chain and undergoes Fabry–Perot–like reflections, leading to maxima in the probability of site occupation, with a periodicity of approximately 220 $w^{-1}$ (we take $\hbar=1$ hereafter)  

Finally, the black curve corresponds to the BIC located at the farthest end, slight initial decay is observed due to leakage in the bulk, consistent with the finite value of the coupling parameter $v$, and subsequent scattering with the bulk states occurs when their Fabry–Perot maxima coincide, giving rise to a small-amplitude beating.

The behavior of the probability as a function of time, depending on the initial condition, can be explained as follows. If the initial condition is that the particle is in the site $j=1$ (red line in Fig.~\ref{fig:rp1}), the closest site to the infinite lead, the state of the system corresponds closely to the symmetrical state; this state is strongly coupled to the continuum. It decays exponentially with a rate $\gamma$, with a beating that takes into account the coupling with the rest of the site\textcolor{blue}{s}  of the SSH chain:
\begin{equation}
P_\mathrm{trivial}(t) = (\alpha \cos^2(wt)+\beta) e^{- \gamma t}\ .
\label{eq:ptrivial}
\end{equation}
where $\alpha$, $\beta$ and $\gamma$ are parameters with the condition, $\alpha+\beta=1$.

On the other hand, the initial condition corresponding to the particle localized in $j=19$ (blue line) shows that the particle feels a trivial chain with loss, coupled to a localized state. The behavior of the return probability can be explained by a phenomenological model, in which we consider a uniform chain with loss in the diagonal elements. Then the probability of finding a particle at the site $j=19$ at $t>0$ is given by the following equation:

\begin{equation}
P_\mathrm{bulk}(t) =  \cos^2(wt) \Big |\sum_ {n = 1}^N c_n e^{-i \varepsilon_n t} e^{-\Gamma_n  t}\Big |^2\ ,
\label{eq:pbulk}
\end{equation}
where $c_n=2/(N+1)\sin[n\pi/(N+1)]$, $\varepsilon_n=-2v \cos[n\pi/(N+1)]$ and $\Gamma_n=[V_0^2/v_c(N+1)]\sin[n\pi (N+1)]$. In this case, the particle is scattered off and partially reflected at the SSH-infinite lead junction. This backscattering effect accounts for the second peak in the probability distribution around $t=220\,w^{-1}$.

If the particle is initially located at the end of the chain, it is almost in an antisymmetric state, indicating that it occupies a BIC. By employing a minimal model with two dimers as a closed system and expanding the states of the system in terms of its eigenstates, we can determine the probability of finding the particle at the site of the chain.
\begin{equation}
P_\mathrm{TBIC}(t)=\frac{1 + 2 r^2 \cos(w t) \cos(v r t) + r^4 \cos(wt)}{(1 + r^2)^2}\ .
\label{eq:ptbic}
\end{equation}
The numerical evaluation of the above equations and the comparison with the results shown in Fig. 4 can be found in the Supplementary Materials.

Figure \ref{fig:rp2} shows the probability density as a function of time and space for the same three initial conditions in Figure \ref{fig:rp1}.  For $j=1$ (red line), the probability decays exponentially in the continuum without coupling with the rest of the chain. In contrast, when the wave packet is initially positioned in the bulk (at j=19, represented by the blue line), it oscillates back and forth due to the finite size of the system. This bouncing behavior explains the non-Markovian nature of the probability. For $ j=20$, the particle remains primarily localized at the end of the chain, exhibiting weak oscillations due to its coupling with the nearest neighbors. As a result, the particle is found in a BIC and does not decay into the continuum.

\begin{figure}
    \centering
 \includegraphics[width=\linewidth]{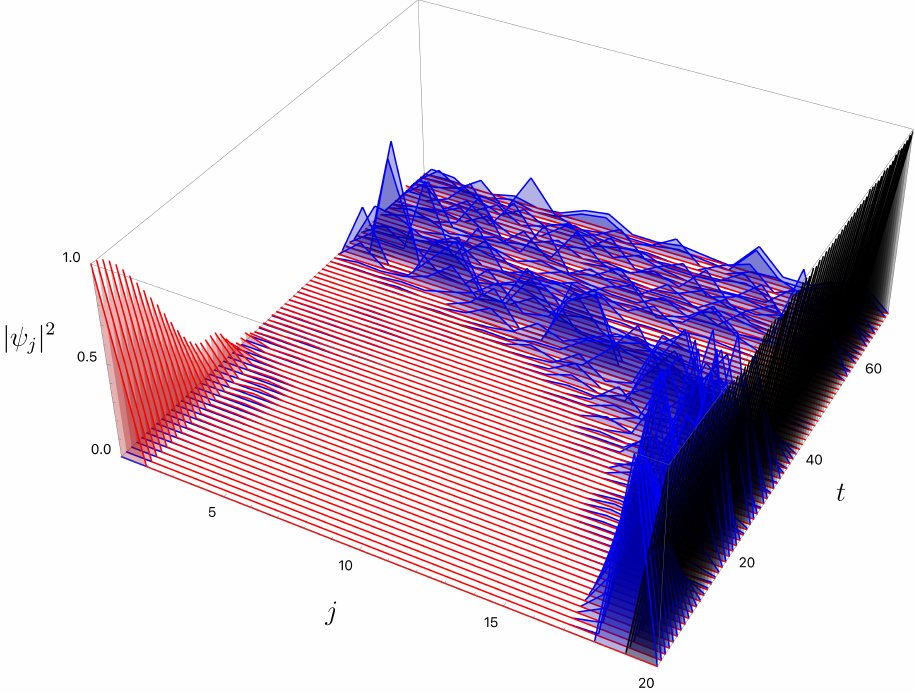}
    \caption{Probability density of two topological states (red and black curves) and one trivial state (blue curve) on a SSH chain of $N=10$ dimer cells attached to the lead, with $\Gamma=0.3$, $J=1$ and $v=0.1$, for three specific sites: $j=1$ (red line), $j=19$ (blue line) and $j=20$ (black line). Time is expressed in units of $w^{-1}$.}
    \label{fig:rp2}
\end{figure}

\textit{Conclusions}--- In summary, we have demonstrated the existence of topologically protected bound states in the continuum (BICs) in a SSH chain. Their robustness arises from a topological obstruction that prevents hybridization with extended modes, ensuring their persistence without fine-tuning. Importantly, we have shown that these states can be unambiguously detected through standard electrical transport measurements, whose dynamic response provides direct evidence of their properties. Our findings offer an experimentally accessible way to identify topologically protected BICs (TBICs) and highlight their significance as stable excitations embedded in the continuum, with important implications for transport and wave-control phenomena.

\begin{acknowledgments}

Work at Madrid has been supported by Comunidad de Madrid (Recovery, Transformation and Resilience Plan) and NextGenerationEU from the European Union (Grant MAD2D-CM-UCM5) and Agencia Estatal de Investigación (Grant PID2022-136285NB-C31/3). P.A.O. acknowledges support from DGIIE USM PI-LIR-24-10, FONDECYT Grants No. 122070, 1230933.

\end{acknowledgments}

\bibliographystyle{apsrev4-2}

\bibliography{references}

\end{document}